\documentstyle[12pt]{article}

\begin{document}

\begin{center}
{\bf \Large Application of the diffraction trace formula to the
three disk scattering system}

Per E. Rosenqvist \\Niels Bohr Institute,
Blegdamsvej 17, DK-2100  Copenhagen \O, Denmark \\
G\'abor Vattay\\
Institute for Solid State Physics, E\"otv\"os University,
M\'uzeum krt. 6-8, H-1088 Budapest, Hungary\\
Andreas Wirzba\\
Institut f\"ur Kernphysik, Technische Hochschule Darmstadt,
Schlo\ss gartenstr.\ 9,\\ D-64289 Darmstadt, Germany

\end{center}
\date{\today}

\pagebreak

\noindent {\bf Abstract:}
The diffraction trace formula
({\em Phys. Rev. Lett.} {\bf 73}, 2304 (1994)) and spectral determinant
are tested on the open three disk
scattering system. 
 The
system contains  a generic and exponentially
growing number of diffraction periodic orbits. In spite of this
it is shown that
even the scattering
resonances with large imaginary part can be reproduced
semiclassicaly.
The non-trivial interplay of the diffraction periodic orbits
with the usual geometrical orbits
produces the fine structure of the complicated spectrum of scattering
resonances,
which are beyond the resolution of the conventional periodic orbit theory.


\pagebreak

\section{Introduction}

Gutzwiller's trace formula$^{\cite{gutz}}$ is an increasingly popular tool for
analyzing semiclassical behaviour. Recently, it has been demonstrated
that using proper mathematical apparatus, like the spectral determinant
of Voros$^{\cite{Voros}}$,
cycle expansions$^{\cite{PCBE}}$ or quantum Fredholm determinants$^{\cite{qfred}}$,
the trace formula can
successfully predict individual eigenenergies of bound systems and
resonances of open scattering systems.
The physical content of the trace formula is the geometrical optical
approximation of quantum mechanics via canonical invariants of closed
classical orbits. This approximation
is very accurate when periodic orbits sufficiently cover the phase space of the
chaotic system.
This is not the case when the number of
obstacles is small or their distance is large compared to their
typical size. Such a problem
occurs where the wave length of a quantum mechanical
(or optical) wave is very large compared to the spatial variation of
a repulsive potential, e.g.\ at the boundaries of
microwave guides, optical fibers, superconducting squids, or circuits in the
ballistic  electron transport,
i.e.\ in most of the devices used for so-called mesoscopic physics.
In such cases it is important to  take into account
the next-to-geometrical effects. In Ref.\cite{GTD}  we have shown how the
{\em Geometric Theory of Diffraction} (GTD)
for hard core potentials can be incorporated in the periodic orbit theory.
We worked out the two disk scattering system as an example, where
diffraction plays an important role.  Since the realization of the importance
of such effects, diffraction periods have been uncovered
in rhomboid billiards$^{\cite{ShiShu}}$, billiards with magnetic
flux lines$^{\cite{Riemann}}$ and in
a limiting case of the hyperbola billiard$^{\cite{Whelan}}$
In the present example we study for the first time a generic
example where an exponential number of diffractive and geometrical orbits
interplay and builds up a complicated spectrum of scattering resonances.
For the reader unfamiliar with the diffraction trace formula
we start by a brief sketch of its
derivation. For a more detailled and completely outlined introduction
to the theory we refer the reader to~\cite{review}.

\section{Diffraction periodic orbits}

As is well known the free particle Green function $G(x,y;E)$ can be
exactly described in terms of geometrical optics by the path that
connects $x$ with $y$ at energy $E$. If a smooth potential is introduced
we have to respond with a refractive index and if hard walls are present
have to deal with diffractive rays to keep the description in the
spirit of geometrical optics.
The diffractive rays connecting two points in the configuration space
can be derived from an extension of Fermat's variational principle
of classical mechanics : Each path connecting $x$ with $y$ has a whole
class of topologically equivalent paths which can all be continuously
deformed into each other without changing the number of encounters
with the hard wall singularities of the system. The generalized
Fermat principle then states that for each such class $\Gamma $ only the
rays of stationary optical length among all the curves in $\Gamma $
contributes to the final field. The total field will
then be the sum of contributions from such
paths with diffractive segments, over all the topologicaly
different classes of paths connecting $x$ with $y$.
%
%
%

Once we know the generalized ray connecting two points
${\cal A}$ and ${\cal B}$
we can compute semiclassicaly the
Green function $G(q_{\cal A},q_{\cal B},E)$ by tracing the
ray$^{\cite{Keller}}$:

{\bf a,} On the geometrical segments of the ray, the Green function
is given by the energy domain Van-Vleck propagator
\begin{equation}
G(q,q',E)=
\frac{2\pi}{(2\pi i \hbar)^{3/2}}D_{\mbox{V}}^{1/2}(q,q',E)
e^{\frac{i}{\hbar}S(q,q',E)-\frac{i}{2}\nu\pi},
\end{equation}
where $D_{\mbox{V}}(q,q',E)=|\det(-\partial^2 S/\partial q_i
\partial q_j')|/|\dot{q}||\dot{q}'|$ is the Van-Vleck determinant
and $\nu$ is the Maslov index (see Ref. \cite{gutz_book} for details).

{\bf b,} When the geometrical ray hits a surface, an edge  or
a vertex of the obstacle it creates a
source for the diffracted
wave. The strength of the source is proportional to
the Green function at the incidence of the ray
\begin{equation}
Q_{\mbox{diff}}=DG_{\mbox{inci}}\ .\label{D}
\end{equation}
The diffraction constant $D$ depends on the local geometry of the obstacle,
 the wave-type and
the nature of the diffraction. It has been determined
in Ref. \cite{Keller} from the asymptotic
semiclassical expansion of the exact solution in a
simple geometry$^{\cite{Keller,Franz}}$. For
the surface diffraction (creeping) it has the form
\begin{equation}
D_l= 2^{{1}/{3}} 3^{{-2}/{3}}\pi e^{5 i {\pi}/{12}}
\frac{(k \rho)^{{1}/{6}}}{Ai'(x_l) }\ .\label{DL}
\end{equation}
Here  $Ai'(x)$ is the derivative of the Airy integral 
$Ai(x)=\int_{0}^{\infty}dt\, \cos(xt-t^3)$,
$k=\sqrt{2mE}/\hbar$
is the wave number, $\rho$ is the
radius of the obstacle at the source of the creeping ray and
$x_l$ are the zeroes of the Airy integral.
The index $l\geq 1$ refers to
the possibility of initiating creeping rays with different modes,
each with its own profile. In practice only the low modes
contribute to the Green function.
For wedge diffraction the diffraction constant is
$$
D=\frac{\sin(\pi/n)}{n}
\left[\frac{1}{\cos(\pi/n)-\cos((\theta-\alpha)/n))}-\frac{1}{
\cos(\pi/n)-\cos((\theta+\alpha+\pi)/n)}\right],
$$
where $(2-n)\pi$ is the angle of the wedge ($n$ is a real number),
$\alpha$ is the incident
angle and $\theta$ is the outgoing angle. For details we
refer to Ref. \cite{Keller}. In the three disk scatterer the edge diffraction
is only important when the system is  closed and we shall
therefore not go deeper into the subject here.

The source, created by the incident ray, then initiates a new ray 
propagating along the surface
(for creeping) or a free ray starting at the edge of the obstacle
(wedge diffraction).

During the creeping  of the ray the amplitude decreases,
which  can be understood as a process analogous to the
radiation processes of electrodynamics.
The radiated intensity is proportional to the intensity of the
ray 
\begin{equation}
\frac{d}{ds}A_l(s,E)^2 =-2  \alpha_l(s,E)\, A_l(s,E)^2,
\end{equation}
where $s$ is the length measured along the surface and $A_l(s,E)$
is the complex amplitude of the Green's function along the surface.
The coefficient $\alpha_l(s,E)$ depends on the local curvature of
the surface, $1/\rho(s)$, and it has the structure
$
\alpha_l(s,E)=x_l e^{-i{\pi}/{6}}({k}/{6\rho(s)^2 })^{{1}/{3}}
$ (see Ref.\cite{SS}),
where the index $l$ refers again to the different modes of the
creeping wave.
The Green's function for the creeping ray of mode $l$ is then given by
\begin{equation}
G^{D}_l(q_{\cal A'},q_{\cal
B'},E)=e^{-\int_0^{L}ds\alpha_l(s,E)}e^{\frac{i}{\hbar}S(q_{\cal
A'},q_{\cal B'},E)},\label{Gcr}
\end{equation}
where $L$ is the length of the arc of the creeping ray,
and $S(q_{\cal A'},q_{\cal B'},E)$ is the  action along it.

When the creeping ray leaves the surface
its intensity can be calculated from the relation (\ref{D})
due to the
reversibility of the Green function.
The total Green function is then the {\em product} of the
Green functions
and diffraction coefficients along the ray.
If for example we have geometrical propagation from
${\cal A}$ to ${\cal A'}$ followed by surface creeping from
${\cal A'}$ to ${\cal B'}$ and then again a geometrical
propagation from ${\cal B'}$ to ${\cal B}$, the total
semiclassical Green function is
\begin{equation}
G(q_{\cal A},q_{\cal B})= G(q_{\cal A},q_{\cal
A'})D_{{\cal A'}}G^{Cr}(q_{\cal A'},q_{\cal
B'})D_{{\cal B'}}  G(q_{\cal B'},q_{\cal B}).
 \label{Gprod}
\end{equation}

Contrary to the pure geometrical case
the  semiclassical energy-domain Green function for rays
with diffraction arcs thus
have a {\em multiplicative} composition law.

When we incorporate diffraction effects into the trace formula,
periodic rays with diffraction segments also contribute.
We can handle separately the pure geometric cycles and the cycles with
at least one diffraction arc or edge:
\begin{equation}
\mbox{Tr}\:  G(E) \approx \mbox{Tr}\:  G_{G}(E) + \mbox{Tr}\:  G_{D}(E),
\end{equation}
where $\mbox{Tr}\:  G_{G}(E)$ is the ordinary Gutzwiller trace formula,
while $\mbox{Tr} \:   G_{D}(E)$
is the new trace formula corresponding
to the non-trivial cycles
of the GTD.
The Gutzwiller trace formula for two dimensional billiards is
\begin{equation}
\mbox{Tr}\:  G_{G}(E)=\frac{1}{i\hbar}\sum_{p}\sum_{r=1}^{\infty}T_p(E)
\frac{e^{irS_p(E)-ir\nu_p\pi/2}}{\mid \Lambda_p^r \mid^{1/2}\left(1-1/\Lambda^r_p\right)},
\end{equation}
where $T_p(E)$ is the time, $S_p(E)$ is
the classical action, $\nu_p$ is the Maslov index and 
$\Lambda_p$ is the stability eigenvalue of the primitive periodic orbit. 
The summation goes for all primitive periodic orbits of the system $p$ and their
repetitions $r$.
$\mbox{Tr}\:  G_{D}(E)$ can be computed by using appropriate
Watson contour integrals$^{\cite{Franz}}$. For technical details we
refer the reader to Refs. \cite{Franz,Wirzba}.
If we denote by $q_i$, $i=1,\dots,n_p$
(with $q_{n_p+i}\equiv q_i$) the
points along
the closed cycle, where the ray  changes from diffraction to pure geometric
evolution or vice versa, or where the ray encounters a wedge diffraction,
the trace for cycles with
{\em at least one diffraction arc}
can be expressed as the product
\begin{equation}
\mbox{Tr}\:  G_{D}(E)=\frac{1}{i\hbar}\sum_p\sum_{r=1}^{\infty}T_p(E)\prod_{i=1}^{n_p}
                              [{D}(q_i)G(q_i,q_{i+1},E)]^r,\label{tracef}
\end{equation}
where $T_p(E)$ is the time period of the primitive cycle
and $D(q_i)$
is the diffraction constant (\ref{DL}) at the point $q_i$.
$G(q_i,q_{i+1},E)$ is
either
the Van-Vleck propagator, if $q_i$ and $q_{i+1}$ are connected by pure geometric
arcs, or is given by the creeping propagator~(\ref{Gcr})
 in case $q_i$ and $q_{i+1}$ are the boundary
points of  a creeping arc. 

\section{The 3-disk system}
To investigate the theory sketched above we apply it to the 3 disk
scattering system.
The 3-disk system has in recent years been subject to a large number
of investigations, and its main virtues are wellknown. Here, we just
would like to remind about some of the basic properties of the system.

The system  consists of three identical disks placed symmetrically
around the origin in a plane (see fig.~\ref{3disksyst}), 
and is completely determined by a single
parameter, namely the ratio $R:a$  of the seperation of the disks
to their radius.
In our calculations we have kept this ratio fixed at $R:a = 6$.

The system thus possesses a
3-fold rotational symmetry around the origin and has 3 reflection
symmetries around the symmetry lines through the center. The system is
therefore invariant under the point group $C_{3v} $; and instead of
considering the system as a whole we can restrict ourselves to the {\em
fundamental domain}$^{\cite{PCBEsym}}$.
 The fundamental domain exactly covers the
whole system when the elements of the point group are applyed to it.
The 3-disk system and a version of the fundamental domain are shown
in figure~\ref{3disksyst}.
For sufficiently large spacing of the disks~$^{\cite{KHpr}}$ the system
has a complete binary symbolic dynamics. All the periodic orbits
can be described in terms of the alphabet $\{ 0,1 \}$ where
`0' corresponds to a bounce under which the particle returns to its starting
disk and `1' corresponds to the bounces where the particle continues
to the next disk. In the fundamental domain there
are  therefore
two fixpoints `1' and `0' corresponding to a triangular,
and back-and-forward bouncing orbit in the full space.
All the geometrical
orbits can be found via a minimization of the path lengths. If
one needs a periodic orbit following a definite sequence of $n$ disk bounces,
one just has to determine the length as a function of the $n$ bouncing positions
and then to minimize this length. That this indeed gives the right periodic
orbit follows from geometrical optics and Fermats principle: when the light
(the particle) follows the shortest path (of a given symbolic sequence),
it will at the same time obey
the reflection law.

The surface of the disk in the fundamental domain can be used as a Poincar\'e
surface of section. Establishing the bouncing map as in Ref.~\cite{pinballpaper}
we can thus calculate the stabilities of the cycles. Following the outlined scheme
we arrive at the results displayed in Table~\ref{geotab}.

A more detailled description of the 3-disk system and 
the methods described in this section 
can be found in e.g. Ref.~\cite{pinballpaper}.

In order to apply the GTD to the calculation of semiclassical resonances, 
we also have to account for the diffraction (creeping)
orbits of the system.
To give an overview of the work to be done, we start by counting the number
of periodic creeping orbits to be evaluated. Because of the symmetry
of the system we can assume that the creeping orbit   always starts tangentially 
from the (half-) disk  in
the fundamental domain which we label disk number 1. 
Considering first an orbit with no geometrical bounces
we see that it has two
different disks to go to, 
and for each  each disk   two different sides to creep
in. This makes a total of four diffraction orbits of topological length 1.
When these
are folded back into the fundamental domain we see that two of them are
self retracing. The two other orbits are  tracing the same
orbit, but in opposite
directions. If we consider paths of the particle with $m$ bounces, 
we see that there will be $2^{n+1}=2^{m+2}$ periodic creeping
orbits of topological order $n$, as for each one of the $m$ bounces the particle
can choose between two disks. 
Thus the number of periodic creeping orbits
grows exponentially fast with the topological length, $n$, of the orbit.
It is quite astonishing however, as we will see later, how few of these orbits are in fact
needed to get a good description of the scattering resonances (including the
ones with large imaginary parts).
The creeping orbits can be described completely by their itinerary
$1 \alpha_1 \alpha_2 \ldots \alpha_n $ where the $\alpha_i $'s are taken
from the alphabet $\{ 1,2,3\} $ and where we do not allow the repeats
$\ldots 11 \ldots$ , $\ldots 22 \ldots$  and  $\ldots 33 \ldots$ .
This description
contains a double  degeneracy due to the fact that the orbit has the choice to
creep around the final disk clockwise or anti-clockwise. 
For instance, `$123$' can
represent two different orbits which start from disk 1 in the fundamental
domain, then hit disk number 2 and finally creep 
around the final disk (3) clockwise or anti-clockwise. 

The restriction that the creeping periodic orbits 
should start and end tangentially on
one of the disks simplifies the search procedure for them considerably:
whereas in the case of geometrical $n$-bounce cycles one
had to minimize a function
of $n$ bouncing parameters, we here only have one parameter in play, namely
the angle where the creeping orbit leaves the initial disk.
Suppose now that we want a specific creeping orbit described by a series
of disk bounces plus the specification of the 
final creeping domain as above. We then scan through
all the angles that leave the first disk  in the fundamental domain. This
gives us an interval of angles where the first wanted disk is being hit.
We then scan this interval for bounces on the next disk in the itinerary
and so on. Finally we scan the last obtained interval to find the angle
under which the ray creeps into the wanted side of the final disk.

\section{Cycle expansion of the spectral determinant}
Having established the data material for the 3-disk system as described
above, we now report on the more technical part of the actual calculation.

The resonances can be recovered from the Gutzwiller-Voros spectral
determinant$^{\cite{Voros}}$ $\Delta(E)$, which is related to the trace formula as
\begin{equation}
\mbox{Tr}\:  G(E)=\frac{d}{dE}\ln \Delta(E).
\end{equation}
The full semiclassical determinant can be written as the
{\em formal}
product of
two spectral determinants, one corresponding to the pure geometrical,
and one to the new diffractional cycles:
$\Delta(E)=\Delta_G(E)\Delta_D(E)$,  due to the additivity of the
traces.
The product is only formal, since the eigenenergies
are not given by the zeros of $\Delta_G(E)$ or $\Delta_D(E)$ individually, but
have to be calculated from a curvature expansion of the {\em combined}
determinant
$\Delta (E)$ itself.

The geometrical part of the spectral determinant is 
given by 
\begin{equation}
\Delta_G(E)=\exp\left(-\sum_{p}\sum_{r=1}^{\infty}\frac{1}{r}
\frac{e^{irS_p(E)-ir\nu_p\pi/2}}{\mid \Lambda_p^r\mid^{1/2}\left(1-1/\Lambda_p^r
\right)}\right),
\end{equation}
where the summations are  over closed primitive (non-repeating)
cycles $p$ and their repetitions  $r$.
The diffraction part of the spectral determinant is
\begin{equation}
\Delta_D(E)=\exp\left(-\sum_{p}\sum_{r=1}^{\infty}\frac{1}{r}\prod_{i=1}^{n_p}
[ D(q^p_i)G(q^p_i,q^p_{i+1},E)]^r\right),
\end{equation}
where the summations are  over closed primitive (non-repeating)
cycles $p$ and their repetitions  $r$.
The product of Green functions should be evaluated for
$q^p_i$ belonging to the primitive cycle $p$.
After summation over $r$, the spectral determinant can be written as
\begin{equation}
\Delta_D(E)=\prod_p(1-t_p) \label{Dp}
\end{equation}
with
\begin{equation}
t_p=\prod_{i=1}^{n_p} { D}(q^p_i)G(q^p_i,q^p_{i+1},E),
\label{We}
\end{equation}
where $q^p_i$ belongs to the primitive cycle $p$.
Here
the mode numbers $l$ of the diffraction constants and the corresponding
summations have been surpressed for notational simplicity;
they can  easily be restored
as e.g.\ in the final expression (\ref{final}).

We can conclude that the  diffractional part $\Delta_D(E)$ of the spectral
determinant shares some nice features of the periodic orbit
expansion of the dynamical zeta functions$^{\cite{CE}}$,
and it can be expanded as
\begin{equation}
\Delta_D(E)=1-\sum_p t_p +\sum_{p,p'}t_p t_{p'}- \cdots .
\end{equation}
The weight (\ref{We}) has the following property  which helps in
radically reducing the number of relevant contributions to the
expansion:
If two different cycles $p$ and $p'$ have at least one common piece in their
diffraction arcs, then the two cycles can be composed to one
longer cycle $p+p'$ and the weight corresponding to this longer cycle
is the product of the weights of the short cycles
\begin{equation}
t_{p+p'}=t_{p}\cdot t_{p'}.
\end{equation}
As a consequence, the product of primitive cycles, which have
at least one common piece in their diffraction arcs, can be
reduced in such a way that the composite cycles are exactly
cancelled in the curvature expansion
\begin{equation}
\prod_p(1-t_p)=1-\sum_b t_b,
\label{creepzeta}
\end{equation}
where $t_b$ are {\em basic} primitive orbits
which can not be composed from shorter primitive
orbits.
In the case of the desymmetrized 3-disk scatterer 
this applies to all the orbits,
and we thus get a zeta function exactly of the form~(\ref{creepzeta}),
where the sum is over all the prime periodic creeping orbits.

To get the free flight part of $ \Delta_D (E) $ we first consider
the semiclassical Green function in free space. This  is asymptotically
($ k R \gg 1$) given as
\begin{equation}
G_0(q,q',E)=-\frac{i}{4}\left(\frac{2}{\pi kR}\right)^{1/2}e^{ikR-i\frac{\pi}{4}
},
\label{GRB}
\end{equation}
where $R=|q-q'|$.
If the ray connecting $q$ and $q'$ is reflected once or more
from the curved hard walls  before hitting tangentially one of the
surfaces, we can keep track of the change in the amplitude by the help of the
Sinai-Bunimovich curvatures.
For a free flight the Sinai-Bunimovich curvature is just the inverse
of the travelled distance
\begin{equation}
        \kappa = \frac{1}{r} .
\end{equation}
When a hard wall is encountered the curvature changes discontinuously as
\begin{equation}
        \kappa_{+} = \kappa_{-} + \frac{2c}{\cos \phi}
\end{equation}
where $\kappa_{\pm}$ are the Sinai-Bunimovich curvatures right after
and before the bounce against the wall and $c$ is the curvature
of the reflecting surface at the point of incidence, whereas $\phi $ is the
angle of incidence.

By computing the curvatures $\kappa_i$ right after the reflections,
and knowing the distances $l_i$ between the $i$-th and the
$(i+1)$-th points of reflections, the factor $R$ in the Green function
(\ref{GRB})
has to be changed to the effective radius
$R^{\mbox{eff}} = l_0\prod_{i=1}^{m}(1+l_i\kappa_i)$ where $l_0$ is the
distance between $q$ and the first point of reflection along the ray as measured
from $q$, and $m$ is the number of reflections from the hard potential walls.
The effective radius $R_b^{\mbox{eff}}$, the length of the geometrical
arc  $L_b^G$ and
the length of the diffraction part $L_b^D$ of the first twelwe orbits with
creeping sections are listed in Table~\ref{creeptable}.
To each cycle in the list, there is a whole sequence
of cycles which wind around the disk $m_{\mbox{w}}$ times. For these
orbits one has to add $2\pi a m_{\mbox{w}}$ to the diffraction length $L^D_b$.
The diffraction part of the spectral determinant is
finally given by
\begin{eqnarray}
\Delta_D(k)&=&1-\sum_{b,l}
(-1)^{m_b}C_l\frac{a^{{1}/{3}}e^{i{\pi}/{12}}e^{ik(L^G_b+L^D_b)-\alpha_l
L^D_b}}{k^{{1}/{6}}\sqrt{R_b^{\mbox{eff}}}} \nonumber \\
& & \times
\frac{1}{1-e^{2\pi (ik-\alpha_l)a}},
 \label{final}
\end{eqnarray}
where $C_l=\pi^{3/2}3^{-4/3}2^{-5/6}/Ai'(x_l)^2$, 
$\alpha_l $ is the
creeping exponent and $m_b$ is
the number of reflections of orbit $b$ from 
the disk in the fundamental domain. The summation
for the windings $m_{\mbox{w}}$ gives the factor
$1/(1-e^{2\pi(ik-\alpha_l)a})$.

\section{Numerical results and conclusions}

To evaluate the results of the diffraction extended Gutzwiller-Voros spectral
determinant, we compare 
the resonances determined by this,
to the resonances determined just from geometrical orbits and
to the exact quantum resonances.

The data are displayed in figures~\ref{res1} and~\ref{res2}. As one
can see the Gutzwiller Voros determinant accounts reasonably well for
the leading order of resonances, whereas it fails for the next
series. In figure~\ref{res2}, however, we can see that -- when a few
periodic creeping orbits are introduced --  the results are
{\em qualitatively different}, and represent  much better  the
trend of the exact quantum resonance data.
For instance,
one can make a one-to-one identification of the
quantum and semiclassical resonances, which is not possible in the purely
geometrical theory, since in that approximation even the number of resonances
is wrong.

The series of subleading resonances also approximately
defines the lower boundary of the region in which the diffractional
spectral determinant still has a high accuracy and good convergence
properties.
This can also be seen from fig.~\ref{res2} since 
for small $\mbox{Re } k$ and large $\mbox{Im } k$
we have a relatively  larger deviation between the exact and creeping
resonances.
   
The errors of the resonances are originated in two sources.
{\bf 1.} The description is semiclassical and therefore we 
use the Van-Vleck propagator in (18) instead of the exact propagator,
and the semiclassical approximation of the creeping propagator in (5).
Also, only the $l=0$ creeping modes are used.
{\bf 2.} Only a restricted number of usual and creeping periodic orbits
is avaliable instead of infinitely many.

As mentioned earlier the number of creeping periodic orbits in this
system increases exponentially with the topological length of the cycles.
It would be natural to expect that this might destroy the simplicity of
the semiclassical description.
We conclude that this seems not to be the case. As we have
demonstrated, one
 only need the basic representatives of the creeping families to change
the picture of the scattering resonances drastically, in the direction
of the exact quantum resonances.

\section{Acknowledgement}

The authors are very grateful to
P. Cvitanovi\'c and P. Sz\'epfalusy for encouragement,
A. Shudo and Y. Shimizu, N. Whelan and S. M. Riemann
for
communicating their results priory to publication.
P.E.R. and G.V. thank the Danish Science Foundation (SNF) for the support
and G.V. acknowledges the support of the Foundation for Hungarian
Higher Education and Research and the Hungarian Science Foundation (OTKA
2090 and F4286).

\pagebreak 
\begin{figure}
\caption{The full 3-disk system with a copy of the fundamental domain.
Representatives
of the creeping orbits of topological length 1 are displayed in full space
as well as
in the fundamental.}
\label{3disksyst}
\end{figure}

\begin{figure}
\caption{The exact quantum mechanical resonances (diamonds) and
the pure  geometrical Gutzwiller Voros resonances (crosses)  in units of $1/a$
in the complex $k$ plane. The resonances belong to 
the one-dimensional $A_1$ representation
of the  3-disk system with $R:a = 6$ .
In
the semiclassical calculation cycles up to topological length 4 has
been used.
The leading resonances close to the real axis are exactly described
by the Gutzwiller Voros resonances
whereas
 the  subleading semiclassical  resonances
 clearly deviates from the exact quantum resonances.}
\label{res1}
\end{figure}

\begin{figure}
\caption{The exact quantum mechanical (diamonds) and
the semiclassical (crosses) $A_1$  resonances of the $R:a = 6$ 
3-disk system. The resonances are calculated by including 
diffractional creeping orbits 
up to order 4
in the GTD. 
 As in the
two disk case
an improvement of the approximation is clearly visible,
especially for the second row of the leading resonances as well as for the
subleading diffractional ones. In the latter case the qualitative trend is
clearly  reproduced. 
As discussed above the accuracy of the semiclassical resonances
becomes worse in the region where 
 $\mbox{Re } k$ is small and  $\mbox{Im } k$ is large.}
\label{res2}
\end{figure}

\pagebreak
\begin{table}
\caption{Geometrical cycle data for the 3-disk system with $R:a = 6$    .    
 The first column
indicates the symbolic dynamics of the periodic orbit, whereas the
second and third column gives the stability calculated from the Jacobian of
the bouncing map, and the length of the cycle 
in the fundamental domain.}
\label{geotab}
\begin{tabular}{r r r c}
$p  $ & $\Lambda_p $ & $ L_p^G/a $                           \\ \hline
 0           &   9.898979     &   4.000000     \\
 1           &  -11.771455     &   4.267949    \\
 10          &  -124.094801    &   8.316529    \\
 100         &  -1240.542557   &   12.321746    \\
 101         &   1449.545074   &   12.580807    \\
 1000        &  -12295.706861  &   16.322276    \\
 1001        &   14459.975919  &   16.585242    \\
 1011        &  -17079.019008  &   16.849071    \\
 10000       &  -121733.838705 &   20.322330    \\
 10001       &   143282.095154 &   20.585689    \\
 10010       &   153925.790742 &   20.638238    \\
 10011       &  -170410.715542 &   20.853571    \\
 10101       &  -179901.947942 &   20.897369    \\
 10111       &   201024.734743 &   21.116994    \\
\end{tabular}

\end{table}

\pagebreak
\begin{table}
\caption{Creeping cycle data for the 3-disk system with 
$R:a = 6$.
 The first column indicates
the itinerary of the orbit, second column the effective radius of the orbit
calculated by means of the Sinai-Bunimovich curvatures and finaly the third
and fourth columns shows the length of the free flight and the creeping
sections respectively.}
\begin{tabular}{r r r c}
$p_c$ & $R_{b}^{\mbox{eff}}/a$ &  $L_b^G/a$   & $L_b^D/a$ \\ \hline
          12 &    6.000000   &  6.000000  &  4.188790   \\
          12 &    5.656854  &   5.656854   & 3.821266 \\
          13 &    6.000000  &   6.000000    & 2.094395 \\
          13 &    5.656854  &   5.656854 & 3.821266 \\
         121 &   58.167840  &   9.832159 &   4.523686   \\
         121 &   58.787753  &   9.797958 &   3.544308   \\
         131 &   58.167840  &   9.832159 &   2.429291   \\
         131 &   58.787753  &   9.797958 &   3.544308   \\
         123 &   66.352162  &   10.120809 &   4.384819   \\
         123 &   73.492203  &   10.147842 &   3.478142   \\
         132 &   84.855171  &   10.120809 &   2.678761   \\
         132 &   73.492203  &   10.147842 &   3.478142   \\
\end{tabular}
\label{creeptable}
\end{table}

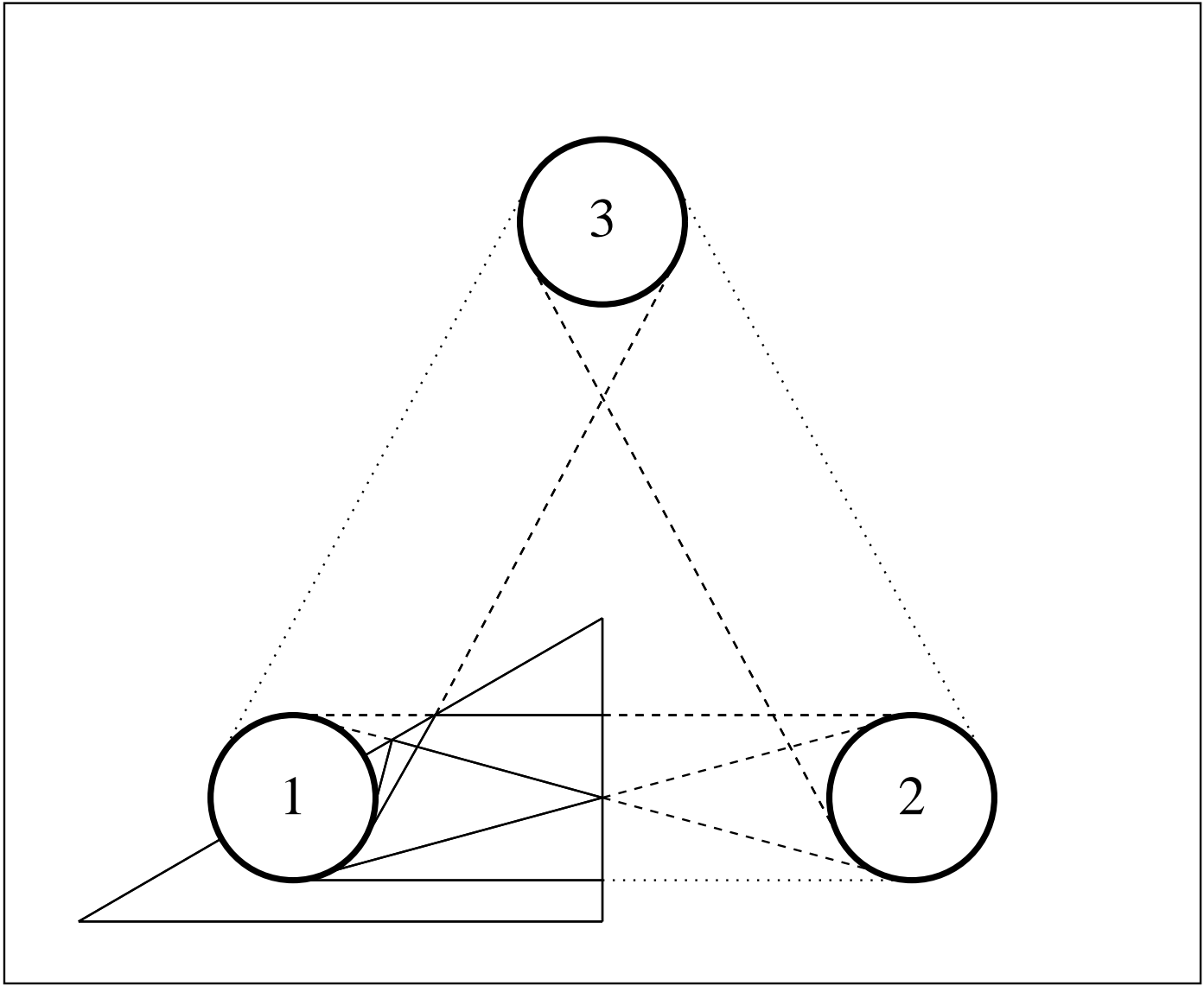
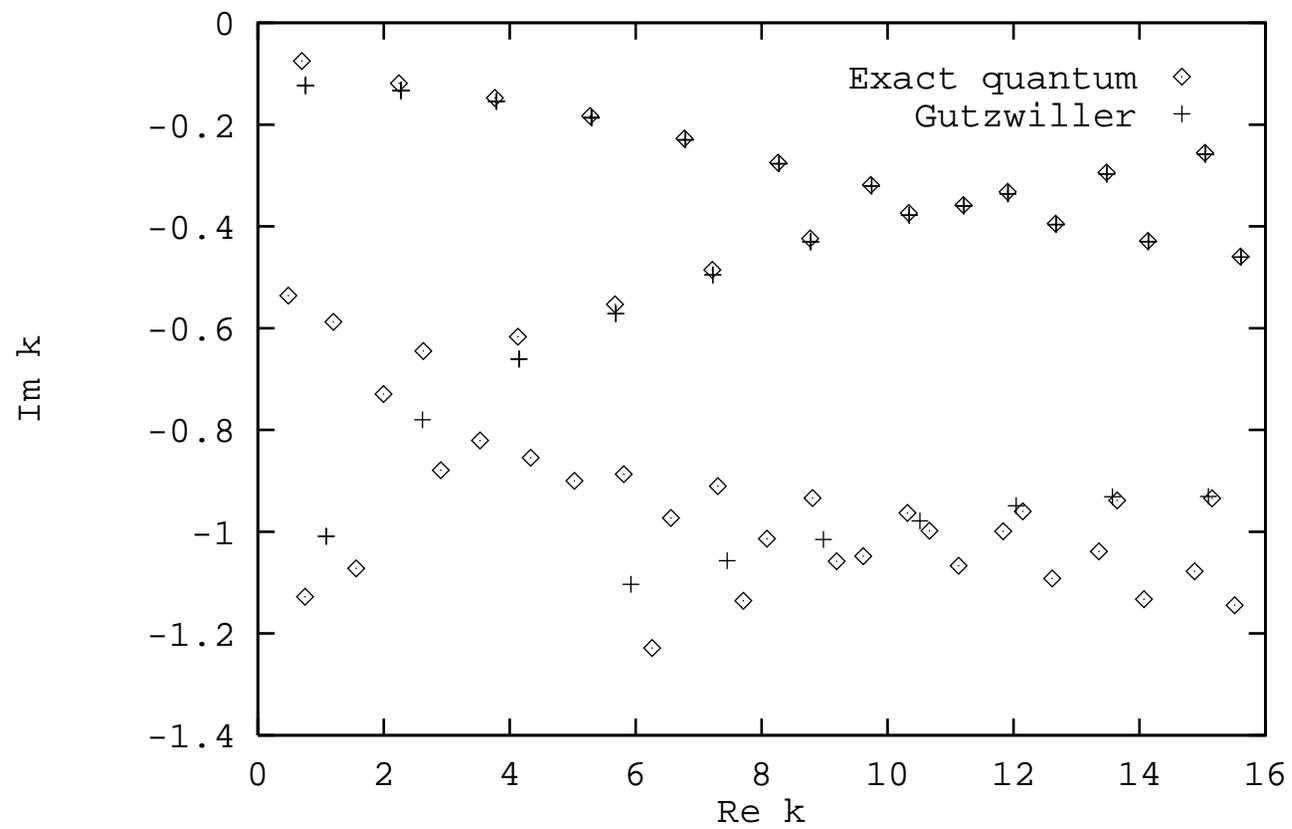
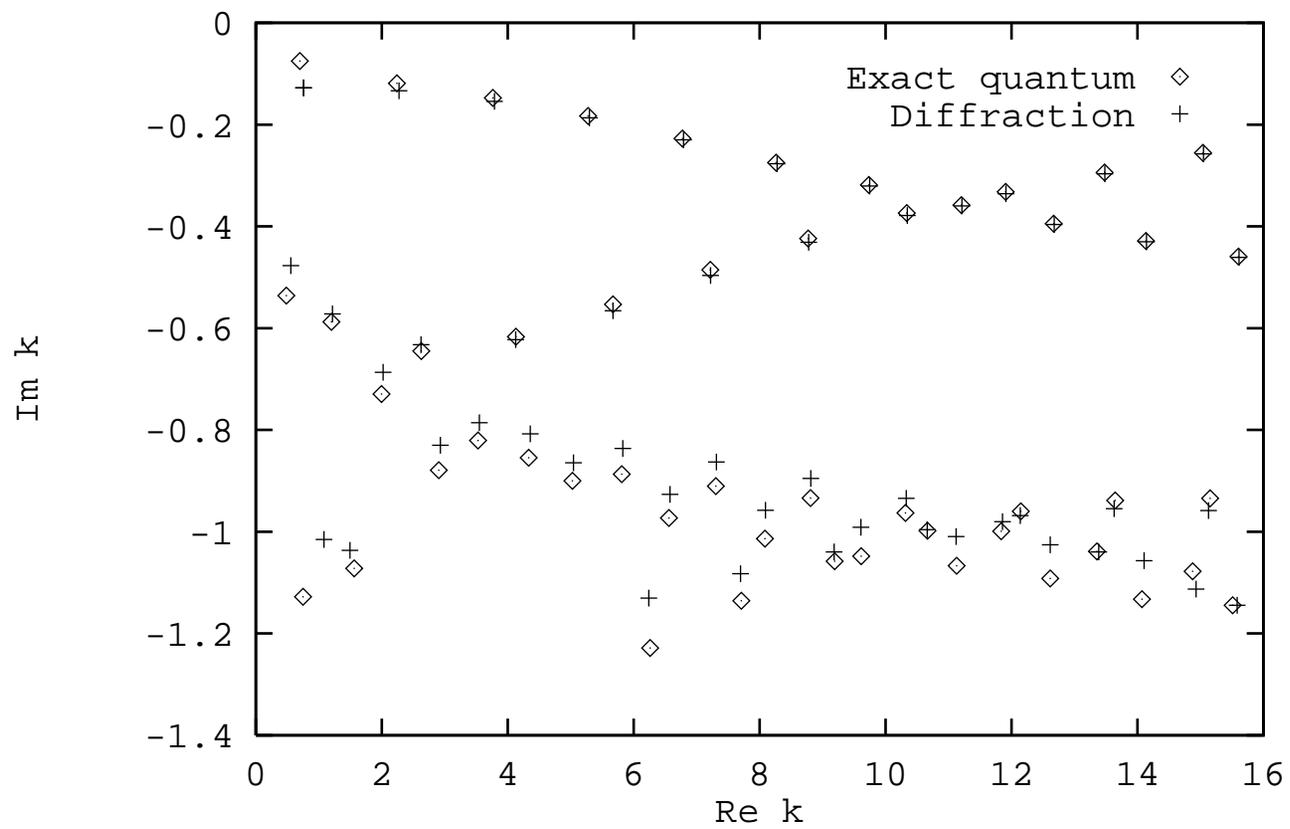
\end{document}